\documentclass[twocolumn]{aastex61}

\shorttitle{Identification of Torsional Waves}
\shortauthors{A.~Chelpanov, N.~Kobanov}
\begin{document}
\title{Problems in Observation and Identification of Torsional Waves in the Lower Solar Atmosphere}
\correspondingauthor{A.~\surname{Chelpanov}}
\email{chelpanov@iszf.irk.ru}
\author{Andrei~\surname{Chelpanov}}
\affil{Institute of Solar-Terrestrial Physics
                     of Siberian Branch of Russian Academy of Sciences, Irkutsk, Russia}
\author{Nikolai~\surname{Kobanov}}
\affiliation{Institute of Solar-Terrestrial Physics
                     of Siberian Branch of Russian Academy of Sciences, Irkutsk, Russia}

\begin{abstract}

Registering periodic variations of spectral line widths serves as the main method for observing torsional Alfv\'{e}n waves.
Theoretically, the method seems valid, yet it entails a number of caveats when applied to data.
For instance, the amplitudes of these observations should vary with changes of the location on the disk, and they should be associated with no intensity oscillations.
We analyze extensive observational material of periodic non-thermal variations of line widths in coronal holes and facular regions in a number of spectral lines: H$\alpha$, He\,\textsc{i} 10830\,\AA, Ca\,\textsc{ii} 8542\,\AA, Ba\,\textsc{ii} 4554\,\AA.
In most cases, we detected associated intensity oscillations at similar frequencies.
Besides, we observed no centre-to-limb dependency.
This calls for a discussion on the practical validity of the method and on the alternative explanations for the nature of non-thermal variations of spectral line widths.
Based on our observations, we consider registering line profile broadening a necessary but not sufficient means for unambiguous identification of torsional Alfv\'{e}n waves in the lower solar atmosphere.
 
\end{abstract}

\section{Introduction} \label{sec:intro}

The input of oscillations and waves in the energy exchange in the solar atmosphere has been discussed for the last seven decades.
\citet{1947MNRAS.107..211A} was the first to suggest that magnetohydrodynamic (MHD) waves might play a role in the solar atmosphere heating.

Alfv\'{e}n waves are the magnetohydrodynamic mode that manifests itself as transverse perturbations of the magnetic field, while the density, temperature, and gas pressure stay constant.
Alfven waves were theoretically predicted several decades ago, and since then, they have been registered and thoroughly observed in the environments where in situ measurements have been available, for example, in the solar wind or magnetosphere of the Earth.

The situation is different in the astronomical studies, where the main -- and sometimes the only -- means of understanding what is going on at the distant objects is analyzing the light and the other types of electromagnetic radiation.
Under this limitation, Alfv\'{e}n waves prove challenging to register, due to their incompressive nature, which allows no possibility to detect them as changes in the brightness of the light emitted by the observed environment or object.

Alfv\'{e}n waves may play a crucial role in the energy exchange processes between the layers of the solar atmosphere.
Some authors consider the torsional Alfv\'{e}n mode dominant in coronal heating \citep{2017ApJ...840...64S,2017NatSR...743147S}.
Researchers also attempt to use observations of torsional waves to map chromospheric magnetic fields \citep{2009Sci...323.1582J,2010ApJ...714.1637V,2011ApJ...733L..15V}.

Registering Alfv\'{e}n torsional waves is theoretically possible studying Doppler-induced variations of spectral line profile widths in non-resolved magnetic tubes.
When a rotating tube is in the aperture of the telescope, its sides emit signals from plasma that moves in the opposite directions along the line of sight.
These divergent Doppler shifts, when not resolved spatially, result in periodically widened spectral line profile.
Note that the greatest input in this broadening comes from the outermost sections of tubes.

The first observations of these variations were carried for coronal lines.
\citet{1959ApJ...130..215B} found that the width of the Fe\,\textsc{xv}  5303\,\AA\ line undergoes periodic variations changing with height.
The author interpreted these oscillations as manifestations of transverse MHD waves.
\citet{1979SoPh...64..223E} also observed the corona in the Fe\,\textsc{xv}  5303\,\AA\ line above an active region.
They found line profile variations with a period of 6 minutes and an amplitude of 35\,m\AA\ alongside with intensity oscillations that were shifted by $180^{\circ}$ compared with the width variations.
No such variations were observed outside active regions in the Fe\,\textsc{xv}  5303\,\AA\ line.
\citet{1973A&A....22..161B, 1975MNRAS.171..697B} studied non-thermal broadening in the 1200\,--\,2200\,\AA\ range in rocket-based data.
Although no direct measurements of line width oscillations were carried out, the authors interpreted the observed broadenings as a result of MHD waves.
\citet{1968SoPh....5....3M} offered to observe Alfv\'{e}n-caused broadenings of spectral lines at different positions of sunspots at the disk of the Sun.
\citet{1978ApJ...226..698M, 1979A&A....73..361M} based on the Skylab observations, concluded that non-thermal broadening of the UV lines may be associated with fast MHD waves.
\citet{1991SoPh..131...41M} proposed observing broadening of non-thermal variations in the transition region under coronal loops while they move across the disk.

The discussion on the nature of the non-thermal broadening of the EUV lines continued the following years.
\citet{1998ApJ...505..957C} and \citet{2008ApJ...673L.219M} showed that line width oscillations in the chromosphere and transition region may be caused in part by unresolved mass flows in spicules.
\citet{2009Sci...323.1582J} gave observational arguments supporting the view that periodic variations of H$\alpha$ line non-thermal broadening result from torsional oscillations in a group of bright points.
The mean amplitude of the observed width oscillations was 56\,m\AA\ in the 126\,--\,700\,s period range with no significant intensity oscillations in this same range.
\citet{2009A&A...501L..15B} studied non-thermal broadening of spectral lines above the limb in the polar regions, which they associated with Alfv\'{e}n waves to the height of 0.1 solar radii.

While this method has often been used for the coronal off-limb observations \citep{2007Sci...318.1574D, 2008A&A...480..509M, 2009ApJ...698..397V, 2012ApJ...751..110B, 2019ASSL..458.....A, 2020ApJ...891...99A, 2021SSRv..217...76B}, signs of torsional oscillations are looked for in the lower solar atmosphere, in search of the sources of the energy that is being transported upwards.
However, the results have been contradictory for the lower atmosphere \citep{2015ApJ...799L..12D}.

In our work, we focus on analyzing a large array of chromospheric observational data to identify non-thermal line width oscillations and to compare them with other observed characteristics that may help to identify them as a result of Alfv\'{e}n waves -- or possibly other processes.
In the analysis, we use objects with different physical characteristics: coronal holes and facular regions.
While the signals registered in the coronal holes are mostly associated with vertical magnetic fields, facular regions harbor low magnetic loops, whose apexes are observed in the chromosphere.
In the case of observing torsional oscillations, this diverse morphology should result in different manifestation of the center-to-limb effect.

\section{Data and Analysis}

We used an extensive data set from the back catalogue of our ground-based spectroscopic observational material that comprises 20 coronal holes and 17 faculae observed over activity Cycles 23 and 24.
The data were collected with the use of the \textit{Horizontal Solar Telescope} at the \textit{Sayan Solar Observatory} \citep{2000SoPh..196..129K,2015SoPh..290..363K}.
The spatial resolution was 1.0\,--\,1.5\,arcsec.
The temporal cadence ranges between 1.5 and 5 seconds.
The observational series lasted between 60 and 180 minutes.
The chosen objects were located in various parts of the solar disk at a wide range of distances from its centre, including the polar regions.
In our observations we used chromospheric lines: H$\alpha$, He\,\textsc{i} 10830\,\AA, Ca\,\textsc{ii} 8542\,\AA, Ba\,\textsc{ii} 4554\,\AA.
The information on the location of the objects at the disk, the spectral lines, the dominant periods in the line-width, intensity, and line-of-sight (LOS) velocity can be found in Tables\,1\,and\,2.

For the analysis we applied the fast Fourier transform algorithm and the Morlet wavelet.
Based on the Fourier spectra, we determined one or two frequency ranges of 1\,mHz width that contain most oscillation power.
The number of frequency ranges was chosen so that they contained 80\% of the power.

\section{Results}

\begin{figure}
\centerline{
\includegraphics[width=6.5cm]{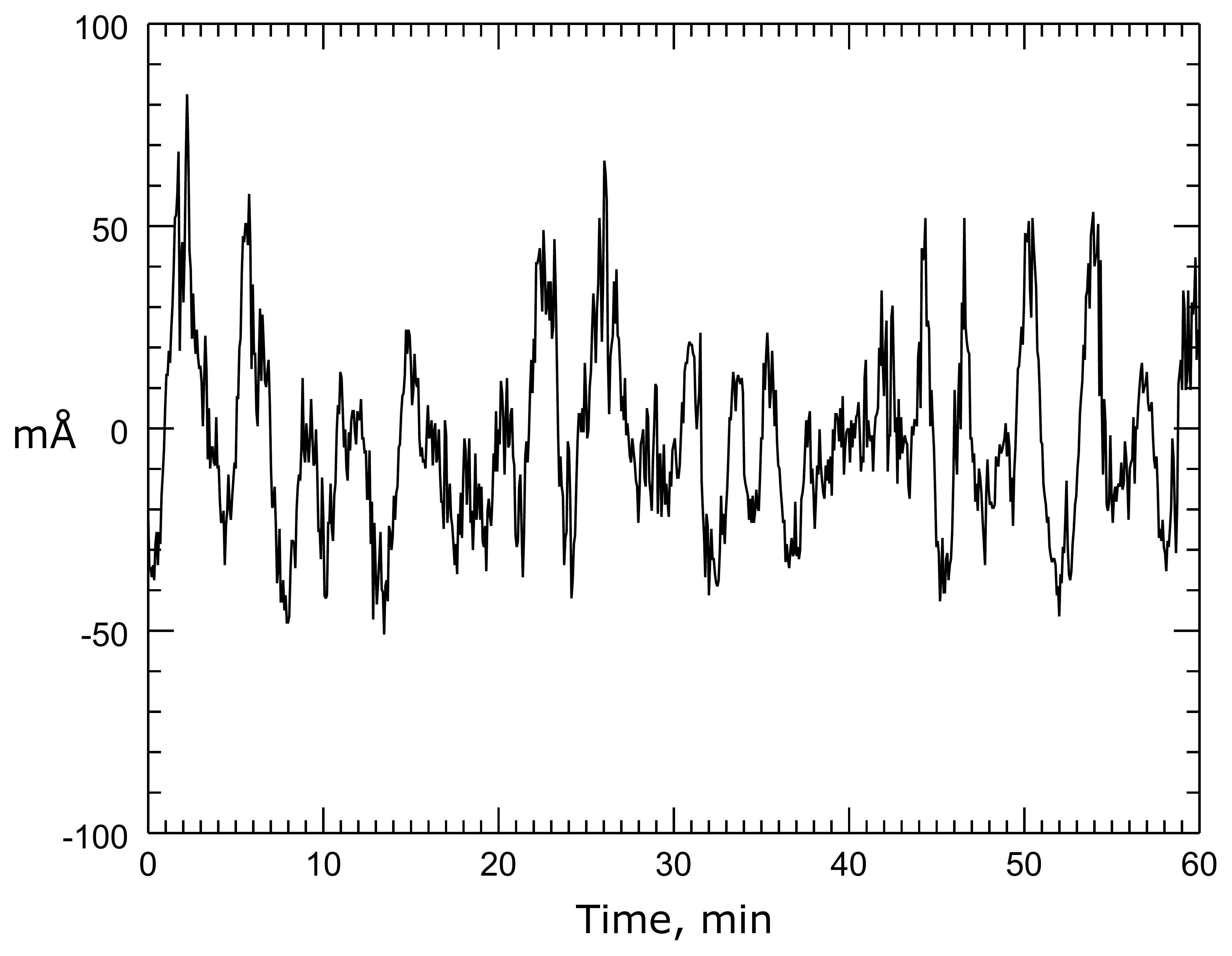}
}
\caption{Example of a profile-width signal of the He\,\textsc{i} 10830\,\AA\ line in a facular region.}
\label{exmpl}
\end{figure}

In all the series, we registered oscillations in the chromospheric line-width signals with amplitudes of tens of m\AA\ up to 100\,m\AA\ (see Figure~\ref{exmpl} for an example of a raw line-width signal).
The exception is the Ba\,\textsc{ii} 4554\,\AA\ line: typical amplitudes in its width variations are 0.5\,--\,2.0\,m\AA.

To compare the observed oscillations with the temperature oscillations that may potentially be registered in these lines, one may use Planck's law, which relates the brightness and temperature:

\begin{equation}
 \epsilon(\lambda, T) = \frac{2 h}{\lambda^5} \frac{c^2}{\exp(\frac{hc}{kT\lambda})-1},
\end{equation}

Based on the observed changes in the intensity of the spectral line and using the temperatures of the line formation, we can calculate the changes in temperature.
From them we are able to derive the changes in the temperature velocities with the use of the following relation:

\begin{equation}
 \upsilon_T = \sqrt{\frac{2kT}{m}},
\end{equation}

The temperature-induced changes in the line width may be estimated from the Doppler-shift formula:

\begin{equation}
 \Delta \lambda = \frac{\upsilon}{c}\lambda,
\end{equation}

In our observational series, all the changes in the brightness fall within 5\% of the background level.
The estimates of temperature-induced changes of the line widths based on these brightness variations yield changes in the line width of three to five\,m\AA, which is 10 to 20 times lower than those we register in the data.

This analysis shows that the oscillations of the line widths that we observe in the chromosphere are of non-thermal origin. Does this mean that they are a product of torsional Alfv\'{e}n waves? Again, we do not have the means to directly measure any parameters to answer this question, but we can try a couple of indirect tests that may help connect these observed oscillations with Alfv\'{e}n waves.

The first detail to address is the small amplitudes in the Ba\,\textsc{ii} line width signals.
The observed amplitudes in this line do not exceed those of thermal origin.
Such a difference from the other lines may be due to the fact that it is attributed to the temperature minimum or the lower chromosphere \citep{1966SvA....10...85S,1983SoPh...82..157M}.
It seems that the more narrow the line, the easier it is to register non-thermal broadening.
These features, however, are only observed in the chromospheric and coronal lines, and the torsional wave signs have not been reported in photospheric lines, although, non-thermal broadening in photospheric lines due to the rotation of the Sun are readily observed in the Sun-as-a-star observations.

\begin{figure}
\centerline{
\includegraphics[width=8cm]{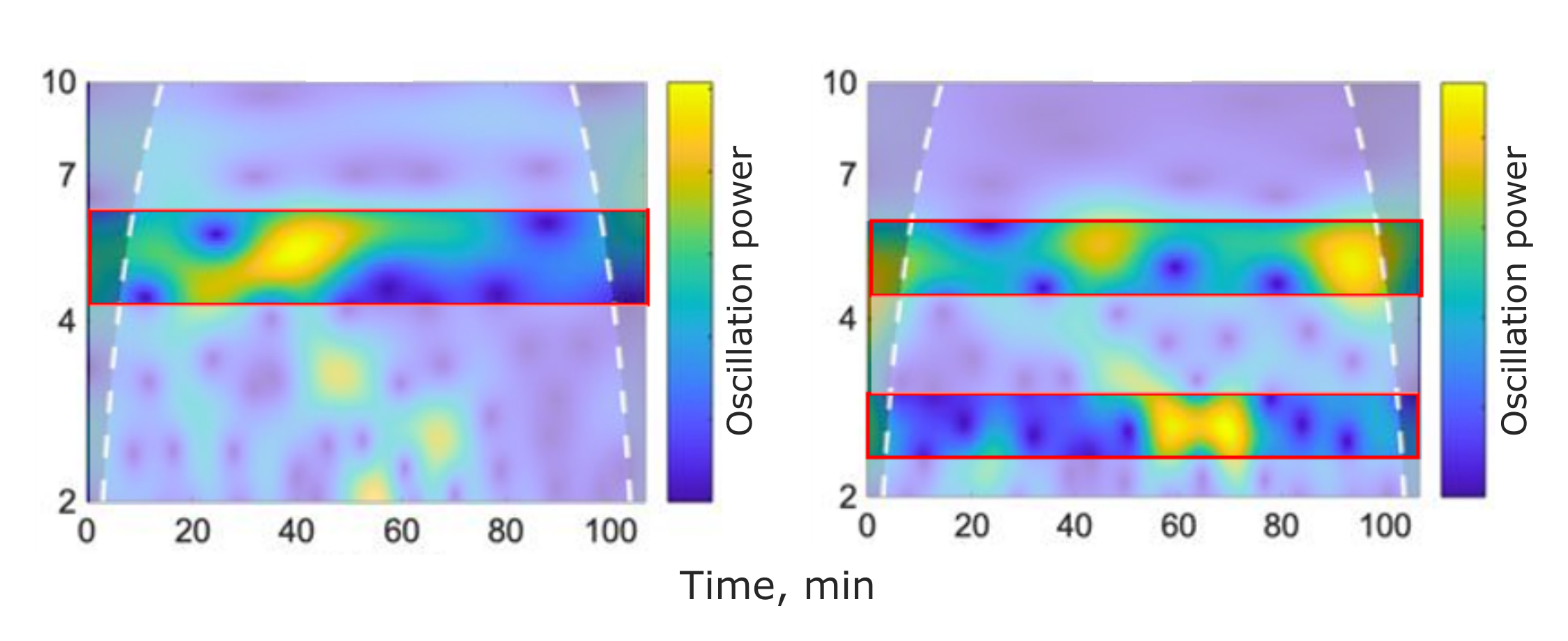}
}
\caption{In the data, we identified the most prominent oscillation periods. In the observed signals, we singled out either one (example in left panel) or two (right panel) ranges of dominant periods. The selected period ranges contain 80\% of the total oscillation power.}
\label{fig:periods}
\end{figure}

\begin{figure}
\centerline{
\includegraphics[width=8cm]{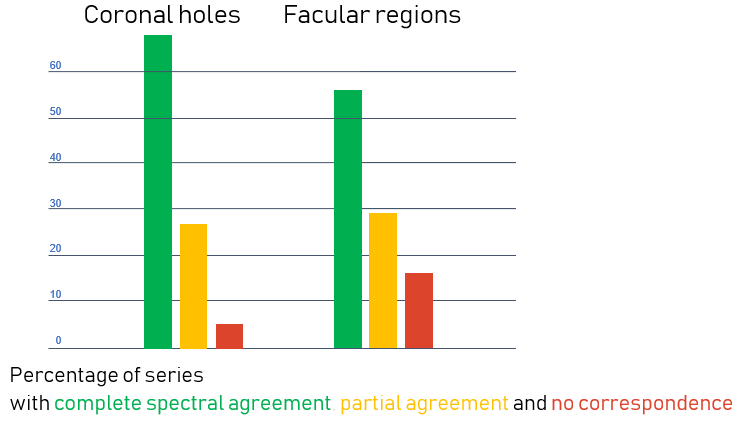}
}
\caption{The percentage of the series showing complete agreement, partial agreement, and no agreement between the spectral composition of the line-width oscillations and intensity oscillations.}
\label{fig:histgrm}
\end{figure}

A way to test the Alfv\'{e}n origin hypothesis would be to compare the intensity and line-of-sight velocity oscillations available in our data with the line-width oscillations.
An absence of apparent similarities between these types of signals might indirectly indicate that the waves responsible for the line-width oscillations are not related with any other MHD mode but the Alfv\'{e}n wave.
We used Fourier analysis and wavelet analysis to single out the most prominent frequencies in signals of line-width variation along with intensity and velocity variations.
For each of the series we identified either one or two frequency ranges that contain most of the oscillation power (Figure~\ref{fig:periods}).
This allowed us to compare the oscillation frequencies in different types of signals.
This comparison shows that most of the series show complete agreement between the dominant frequencies in the line-width and intensity oscillations (Figure~\ref{fig:histgrm}).
As for the rest of the series, the dominant frequencies coincide in more than a half of the cases.

Thus, for most of the objects that we observed, the line-width oscillations are accompanied by oscillations of similar frequencies in the signals of the other parameters, which contradicts the Alfv\'{e}n mode assumption.

\begin{figure}
\centerline{
\includegraphics[width=8cm]{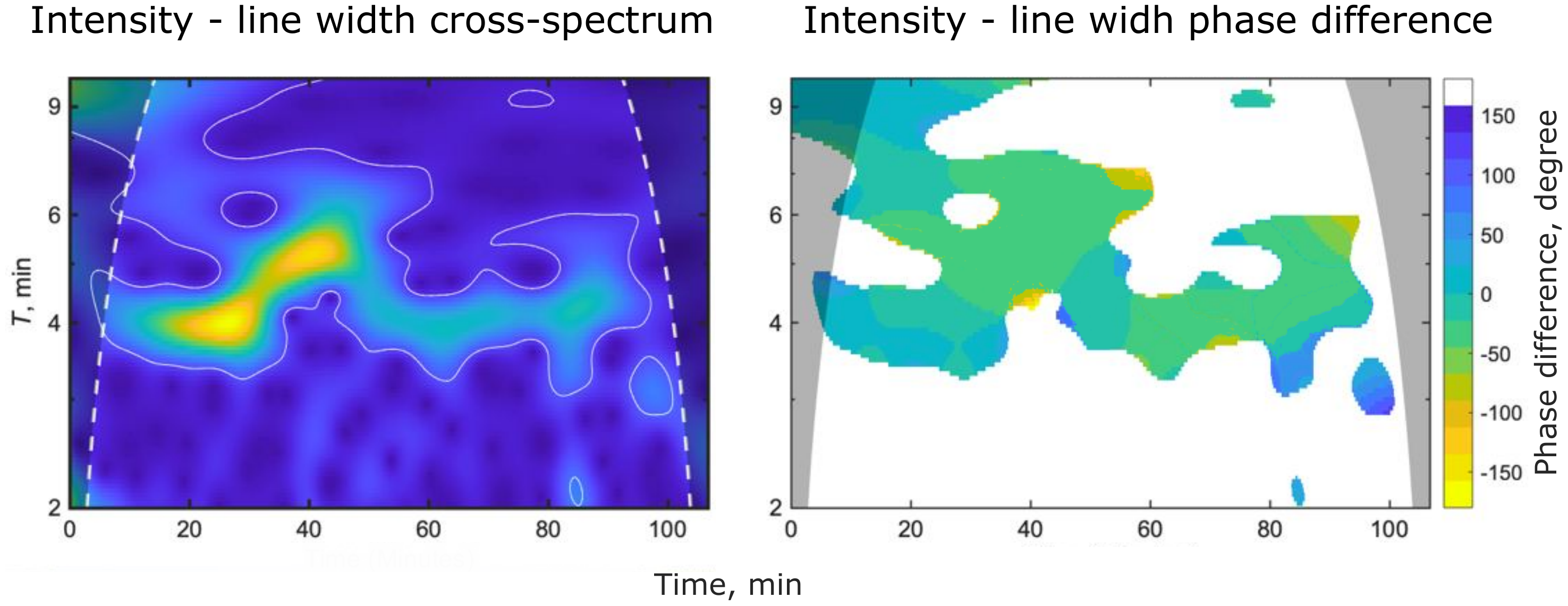}
}
\caption{\textit{Left}: An example of wavelet cross-spectra for the intensity and line-width signals, which indicates periods and times where both signals show significant oscillation power. \textit{Right}: The phase difference between oscillations in the two signals shows little variation.}
\label{fig:phase}
\end{figure}

We also analysed the phase difference between the line-width oscillations and intensity oscillations (Figure~\ref{fig:phase}).
For the dominant frequency-oscillations the difference in the phase stays stable throughout the observation period, which further suggests the relation between these two parameters.

Another way to test whether the observed chromosphere profile-width oscillations are a product of Alfv\'{e}n waves is to track the dependence of their amplitudes on the location at the disk of the Sun \citep{1991SoPh..131...41M}.
Due to its transverse nature, a torsional Alfv\'{e}n wave should manifest itself in the profile width signals the most when a magnetic tube is observed from a side perpendicularly.
In the case when a tube is parallel to the line of sight, Alfv\'{e}n waves should give no input into the signals.
Assuming that the magnetic tubes in the chromosphere are at least somewhat oriented vertically, we might expect to observe higher amplitudes in the profile width oscillations closer to the limb positions, and lower amplitudes towards the centre of the disk.

\begin{figure}
\centerline{
\includegraphics[width=9cm]{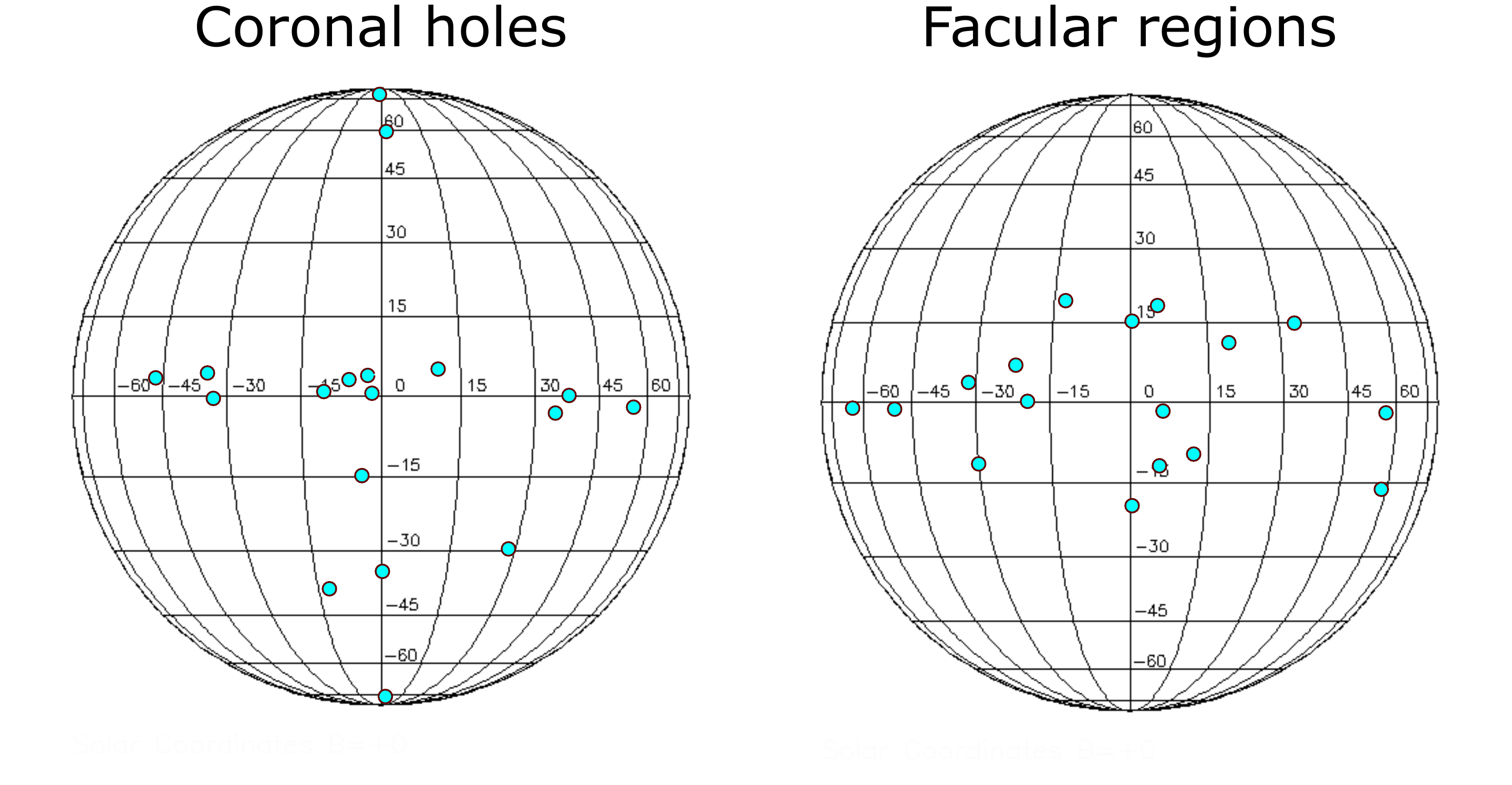}
}
\caption{The locations of the observed objects on the solar disk. The data cover a wide range of distances from the center.}
\label{fig:locs}
\end{figure}

The observation series that we used cover a wide range of locations on the disk from its center to the polar regions, which provides different observational angles (Figure~\ref{fig:locs}).

\begin{figure}
\centerline{
\includegraphics[width=8cm]{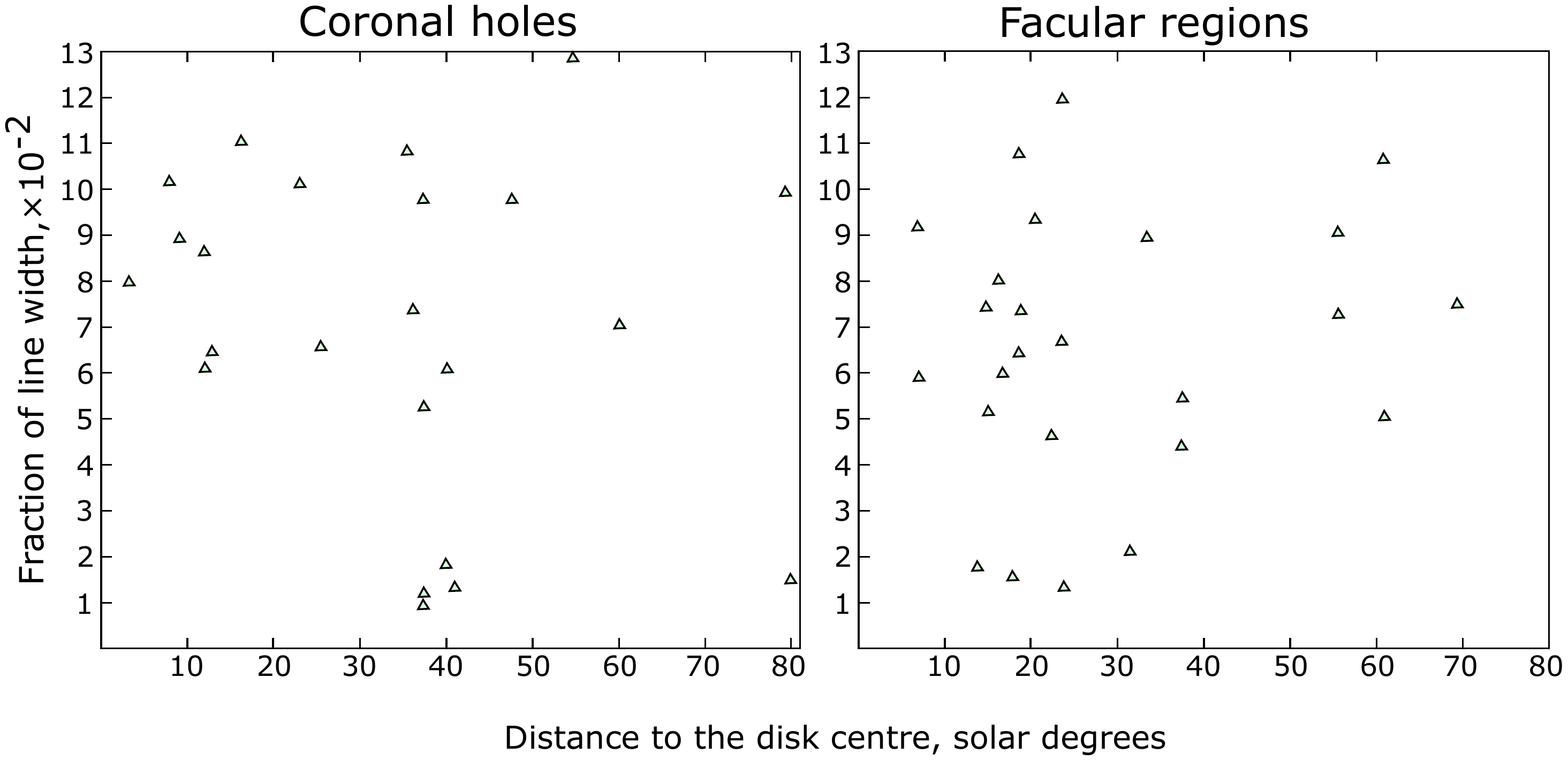}
}
\caption{Relative amplitudes (amplitudes divided by the line width averaged over the time series) of the line-width oscillations plotted against the center-limb locations. The diagrams show no dependence on the observational angle.}
\label{fig:diagrm}
\end{figure}

The data, however, show no dependence of the observed amplitudes.
For both coronal holes and faculae, the amplitudes range between 5 and 100\,m\AA\ both around the disk centre and in the regions where the angle between the line of sight and the normal to the surface reaches 60\,--\,$80^{\circ}$ (Figure~\ref{fig:diagrm}).
Thus, these distributions in our data do not support the Alfv\'{e}n-based explanation of the profile-width variations.

\section{Discussion}

Our observations show that periodic chromospheric line broadenings of a non-thermal nature are ubiquitous in various types of objects.
Earlier, similar results were shown based on observations with the same telescope \citep{2014AstL...40..222Z,2016AstL...42...55K}.
These oscillations show no dependence on the observation angle;
they often coexist with oscillations of the same frequencies in the intensity signals and (to a lesser extent) line-of-sight velocity signals.
This undermines the reliability of registering Alfv\'{e}n waves using this method and calls for a discussion on the alternative explanations for the nature of non-thermal variations of spectral line widths.

The lack of a dependence on the observation angle has also been shown in \citet{1998A&A...337..287E} and \citet{2015ApJ...799L..12D}.
It may be explained by the changing optical depth closer to the limb, and as a result, the line of sight passes through multiple elements in different physical conditions.
A phase and frequency mismatch in these elements may lead to an increase in the constant component of the non-thermal broadening and a decrease in relative amplitude of the periodic variations.

It has been suggested that the Doppler broadening of the lines is due to several loops containing slow waves at different phases that superimpose in the line of sight \citep{1959ApJ...130..215B, 2003A&A...399L..15Z}.
Such an explanation could be consistent with a high correlation between the half-width and brightness oscillation spectra.

Propagation in a non-uniform environment, for example, bent magnetic tubes, may also explain the possible manifestation of torsional waves in the intensity signals.
It has also been shown that Alfvén waves induce longitudinal plasma motions and density perturbations \citep{2011A&A...526A..80V,2022MNRAS.515.5151B}.

Note that even in the absence of intensity oscillations associated with periodic line-width variations, they should not be considered as a unique and sufficient indication for a definite identification of torsional oscillations. For example, a similar situation may take place when observing sausage modes with a limited spatial resolution \citep{2012A&A...543A..12G}.
However, as \citet{2013A&A...555A..74A} and \citet{2017AstL...43..844K} showed, the intensity oscillations in this case should be observed at the half-frequency relative to the frequency of the line-width oscillations, which is not the case in our data.
Thus, the sausage-mode explanation seems improbable for our observations.

\citet{2015ApJ...799L..12D} studying oscillations in the transition region suggested that non-thermal broadening of the line profiles may result from shock waves that have an acoustic origin.
These shocks leaking from the photosphere cause spectral line broadening due to an increased range of velocities before and after the passage of the shock.
This mechanism may explain the connection between the line width and intensity variations.
The effect of the shocks in the data should manifest itself better in the regions, where the magnetic field is aligned with the line of sight, i.e., in the central parts of the solar disk.
In our data, however, we observe no dependence of the line width oscillation amplitude on the distance from the center.
Also, these shocks may not be a coherent explanation for the oscillations that we observe, since they appear irregularly, while the oscillations in our observations are stable with a constant phase.

Localized chromospheric swirls \citep{2009A&A...507L...9W} may induce line broadening too, since they also represent small-scale rotating structures. \citet{2021A&A...649A.121B} directly associate them with Alfv\'{e}n waves. But they appear sporadically, as shock waves do, which rules out this scenario for our observations.

A better perspective might be gained from observing Stokes V profiles in the magnetic elements.
This approach may single out the input of magnetic elements from the total registered radiation.
However, one should keep in mind that the efficiency of this method is subdued by the fact that the Doppler-shift effect is maximal for the outer parts of the tube where the polarization signal is weaker.

As for the future of studying torsional Alfv\'{e}n waves, we may expect more robust results from observing rotations directly with high-resolution instruments.
This direct method, though, poses certain constraints as well, since it requires sufficiently contrast objects whose shape stands out against the background.
Another direct method was proposed by \citet{2017NatSR...743147S}, who reported observations of spatially resolved opposing Doppler shifts in the two halves of magnetic flux tubes, although the duration of the oscillations that they registered does not exceed one period.

\section{Conclusions}

Non-thermal broadenings of spectral lines in the chromosphere are ubiquitous, and they theoretically may be used to register torsional Alfv\'{e}n waves.
The observations, however, are complicated for interpretation.
On the one hand, spatially unresolved rotations undoubtedly cause non-thermal broadenings of spectral lines.
In observations, however, these broadenings are often accompanied with intensity oscillations of the same frequency, which should not be the case for Alfv\'{e}n waves.

We consider several possible explanations. One of them offered by \citet{2015ApJ...799L..12D} suggests that the broadenings are due to acoustic shock waves.
The other one implies different types of waves existing in the same volume.
For example, torsional Alfv\'{e}n waves may induce longitudinal magnetoacoustic waves \citep{2011A&A...526A..80V,2022MNRAS.515.5151B}.

Based on our observations, we consider registering line-profile broadening a necessary but not sufficient means for unambiguous identification of torsional Alfv\'{e}n waves in the lower solar atmosphere.

\acknowledgments

The selection and reduction of the data and scientific analysis were supported by the Russian Science Foundation under Grant 21-72-10139.
The spectral data were obtained using the equipment of Center for Common Use \textit{`Angara'} http://ckp-rf.ru/ckp/3056/ with financial support by the Ministry of Science and Higher Education of the Russian Federation.

\bibliography{Chelpanov}

\end{document}